\documentstyle[12pt,epsf]{article}
\begin{document}

\title{\bf MC-Simulation of the\\ Transverse Double Spin Asymmetry for
RHIC\footnote{UFTP 417 preprint/96}} 
\author{Oliver Martin, Andreas Sch\"afer\\
\\
Institut f\"ur Theoretische Physik\\
J.W.~Goethe--Universit\"at Frankfurt\\
Germany}
\thispagestyle{empty}
\maketitle

\parindent0cm
\begin{abstract}
Using {\sc Sphinx tt}, a new MC simulation program for transverse polarized 
nucleon--nucleon scattering based on {\sc Pythia~5.6}, 
we calculate the transverse double spin asymmetry $A^{TT}$ 
in the Drell-Yan process. If one assumes (quite arbitrarily)
that the transversity parton distribution $\delta q(x,Q^2)$ 
equals the helicity distribution $\Delta q(x,Q^2)$ at some low
$Q_0^2$ scale, the resulting asymmetry is of order 1\%.
In this case is $A^{TT}$
would hardly be be measurable with PHENIX at RHIC. 
\end{abstract}
\parindent1cm
\newpage

\section{Introduction}
Originally, the transversity parton distribution 
$\delta q(x,Q^2)$\footnote{Other common denotions for 
$\delta q(x,Q^2)$ are $h_1^q(x,Q^2)$, $h_T^q(x,Q^2)$, and $\Delta_1 q(x,q^2)$.} was 
defined by
Ralston and Soper \cite{ralston79}.
However, only after polarized targets were developped and first
spin-experiments
yielded very interesting results, transverse spin structure functions
became a real issue.
After the ``rediscovery'' of $\delta q(x,Q^2)$ by Artru and Mekhfi 
\cite{artru90} in 1990 the topic of 
transverse spin has gained an increasing amount of attention by elementary particle
physicists. Much of the interest is caused by the fact that
besides the well known unpolarized and helicity weighted parton distributions
$q(x,Q^2)$ and $\Delta q(x,Q^2)$ the transversity distribution 
$\delta q(x,Q^2)$ is the only
twist-2 distribution of the nucleon. 
On the other hand, its special chirality selection rules
prevent its measurement at leading order in polarized {\em inclusive} deep 
inelastic
lepton--proton scattering experiments and therefore no experimental
data on $\delta q(x,Q^2)$ is available up to now. This situation is
expected to change in the near future, when  the Relativistic Heavy Ion 
Collider (RHIC) at Brookhaven National Laboratory will be
able to scatter transversely polarized protons with high cm energy,
luminosity, and large polarization of each beam. One of the proposed 
experiments
for the RHIC spin program is the measurement of the transverse 
double spin asymmetry 
\begin{equation}
A^{TT}=\frac{{\rm d}\sigma^{\uparrow\uparrow} 
-{\rm d}\sigma^{\uparrow\downarrow}}
{{\rm d} \sigma^{\uparrow\uparrow} +
{\rm d}\sigma^{\uparrow\downarrow}} 
\end{equation}
using the Drell--Yan process
$q_i\bar q_i \rightarrow l^+l^-$ 
\cite{rsc92,rsc93,rsc94},
where $i$ denotes the flavour and 
$l^+l^-$ an electron-positron or muon pair. 
Here, the asymmetry is not suppressed because the
two quark lines in the unitarity diagram are independent of each other
\cite{artru90}. This also holds for other partonic 
processes, like the prompt photon 
production with an away-side jet $q_i + \bar q_i \rightarrow \gamma g$, 
or two-jet production $q_i + \bar q_i \rightarrow gg$,
$q_i q_i \rightarrow q_i q_i$. Nonetheless, 
these reactions 
are not very useful for extracting information about the transversity
distribution, since either the partonic part of the asymmetry is small
or the cross section of other contributing background processes is much larger
\cite{ji92,jaf96}. 

It is generally believed, that at some low scale $Q_0$ the 
transversity distribution is similar to the helicity distribution 
\begin{equation}
\label{assumption}
\delta q(x,Q_0^2)\simeq \Delta q(x,Q_0^2)
\end{equation} 
and that
their difference is only due to dynamic effects \cite{jaf92,he95}.
Crude analytical estimates which were based on this assumption yielded
$A^{TT}\sim 10\%$ \cite{ji92,sof95}. However, in these model calculations the
correct $Q^2$--evolution of $\delta q(x,Q^2)$ was not taken into account.  

In this paper we present a new estimate of the transverse double spin
asymmetry of the Drell-Yan process based on (\ref{assumption})
using the new MC simulation program {\sc Sphinx tt}. We take into account
the correct $Q^2$--evolution and give a crude estimate for the
QCD corrections to the simple Drell-Yan model. 

In section~\ref{2} we describe the model we used for our calculations.
Section~\ref{3} contains a short description  
of {\sc Sphinx tt}. Our results and conclusions are presented in
section~\ref{4}.

\section{Our model \label{2}}
In order to be able to make an estimate of $A^{TT}$ we assume
that the transversity weighted distribution of quarks and antiquarks equals the
helicity weighted distribution at the low scale  
$Q_0^2=4$~GeV$^2$:
\begin{equation}
\delta q(x,4\;{\rm GeV}^2)=\Delta q(x,4\;{\rm GeV}^2),\quad  
\delta \bar q(x,4\;{\rm GeV}^2)=\Delta \bar q(x,4\;{\rm GeV}^2)\quad .
\end{equation}
This is rather an ad~hoc assumption. Transversity distributions and
helicity distributions should agree for some very small $Q^2$, where
perturbative QCD is not applicable. It is not clear how much they will
differ after evolution up to 4~GeV$^2$. Their DGLAP equations differ
most notably because the transversity distribution is not a singlett
with respect to chirality and therefore does not couple to gluons. Its
evolution equation reads simply 
\begin{equation}
\label{glap}
\frac{{\rm d}}{{\rm d}\ln Q^2}\delta q_i(x,Q^2) 
=\frac{\alpha_s(Q^2)}{2\pi} \int_x^1 \frac{{\rm d}y}{y}
\delta q_i(y,Q^2)\;\delta P_{qq} \left(\frac{x}{y}\right)\quad .
\end{equation} 
The splitting function $\delta P_{qq}(z)= C_2(R)[2/(1-z)_+ -2
+ (3/2)\delta(z-1)]$ was calculated by Artru and Mekhfi \cite{artru90}.
Equation (\ref{glap}) implies for the anomalous dimensions

\begin{eqnarray}
\delta q^{(n)}_i(Q^2) &=& \int_0^1 
{\rm d}x\; x^{n-1}\delta q_i(x,Q^2)\quad,
\\
\delta P_{qq}^{(n)} &=&\int_0^1 {\rm d z}\;z^{n-1}\delta P_{qq}(z)
=C_2(R) 
\left(\frac{3}{2}-2\sum_{j=1}^n \frac{1}{j} \right)\quad ,
\\
\delta q_i^{(n)}(Q^2)&=&\delta q_i^{(n)}(Q_0^2) 
\left(\frac{\alpha_s(Q_0^2)}
{\alpha_s(Q^2)}\right)^{\delta P^{(n)}_{qq}/2\pi b}\quad .
\end{eqnarray}  

Since $b=(33-2n_f)/12\pi > 0$, $\delta P_{qq}^{(n)}<0$ and  
$\delta P_{qq}^{(n+1)}<\delta P_{qq}^{(n)}$ all momenta of the transversity
distribution decrease and the distribution becomes softer with increasing
$Q^2$.

We used the leading-order para\-metrizations Set~C of Gehr\-mann/Stir\-ling 
\cite{gehr95}, and  
Bartelski/Tatur \cite{bar95} and evolved them according to (\ref{glap}).

The authors of all major previous works \cite{ralston79,artru90,ji92,jaf92,
cortes91} about the transverse double spin asymmetry of 
lepton pair production only considered the simple electromagnetic 
graph $q \bar q\rightarrow l^+l^-$ which yields the hadronic double
spin asymmetry
\begin{equation}
\label{asydy}
A^{TT}_{DY} = \frac{\sin^2 \hat \theta \cos 2\hat \phi}{1+\cos^2 \hat \theta}
\;\frac{\sum_i e_i^2 \delta q_i(x_A,Q^2)\, \delta \bar q_i(x_B,Q^2) + 
A\leftrightarrow B}
{\sum_i e_i^2 q_i(x_A,Q^2)\, \bar q_i(x_B,Q^2) + A \leftrightarrow B}\quad .
\end{equation}
$\hat \theta$ and $\hat \phi$ are the polar and azimuthal angles
of the lepton momentum in the dilepton rest frame
with respect to the beam and spin axis, resp. 

However, it is well known that the cross section
of the QCD corrections at ${\cal O}(\alpha_s)$ is approximately
as large as the one of the Drell-Yan graph resulting in a 
$K$-factor of
\begin{equation}
K^{DY}=\frac{{\rm d}\sigma_{{\cal O}(\alpha_s^0)} + 
{\rm d}\sigma_{{\cal O}(\alpha_s)}}
{{\rm d}\sigma_{{\cal O}(\alpha_s^0)}}\approx 2\quad .
\end{equation}  
Since this Bremsstrahlung graphs might carry asymmetries completely
different to (\ref{asydy}) one should have a closer look at them.
In general, one would expect the analytical expressions for these
asymmetries to be more complicated than (\ref{asydy}) since one more
outgoing parton is involved here. So it would be nice, if one could
seperate the Drell-Yan graph from its QCD corrections kinematically.

\begin{figure}[hbt]
\begin{center}
\epsfysize=4cm
\leavevmode\epsffile{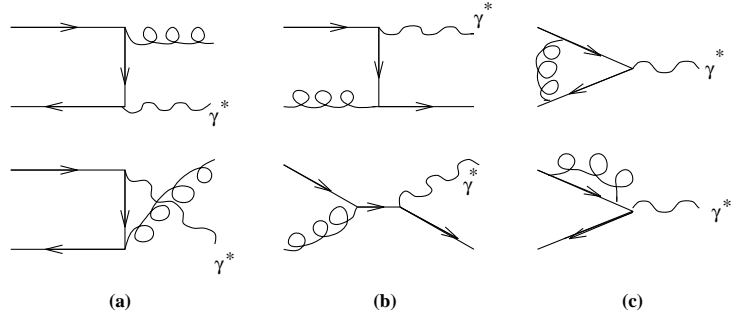}
\caption{\label{fig1}QCD corrections at order $\alpha_s$ 
to the simple Drell-Yan model: (a) annihilation graph (b) Compton graph,
(c) virtual corrections. The decay of the virtual photon $\gamma^*$ is not
shown.}
\end{center}
\end{figure}

The transverse momentum $q_T$ of the virtual photon produced in the Drell-Yan
model is of the order of 1 GeV since the quark and antiquark possess
a primordial transverse momentum in the proton of the order of 
several hundred MeV. Contrarily, the virtual photon in the QCD-Compton
and QCD-annihilation process can also posses a large transverse momentum,
which is balanced by the large opposite transverse momentum of the 
additional outgoing parton. In case the transverse momentum of the virtual 
photon in the QCD annihilation or Compton process is small, the 
reaction is characterized by three scales, namely $q^2$, $q_T^2$, 
$\Lambda_{\rm QCD}^2$.
Here, the differential cross section is dominated by  
double-leading-logarithmic contributions
\begin{equation}
\label{lla}
\frac{{\rm d}\sigma}{{\rm d}q_T^2\,{\rm d}q^2} \propto
\frac{\alpha_s(q^2)}{q_T^2} \ln\left(\frac{q_T^2}{q^2}\right)
\left[b_1+b_2\alpha_s(q^2)\ln^2\left(\frac{q_T^2}{q^2}\right)
+b_3 \alpha^2_s(q^2)\ln^4\left(\frac{q_T^2}{q^2}\right)+\dots \right] 
\end{equation}
where the $b_i$ are constants.
The true expansion parameter is rather 
$\alpha_s(q^2)\ln^2(q_T^2/q^2)$ than $\alpha_s(q^2)$.
Equation (\ref{lla}) describes the radiation of an infinite number of
soft gluons from the partons which enter the hard partonic reaction. 
Fortunately the coefficients $b_i$ are not independent of each other but
exponentiate \cite{field}.

The differential cross section ${\rm d}\sigma/{\rm d}q_T^2{\rm d}q^2$ 
at double-logarithmic precision  differs from the one 
calculated by a naive convolution only by a form factor 
$T_F^2$ for the QCD-annihilation process and $T_FT_G$  for the
QCD-Compton process where \cite{dok80}
\begin{equation}
\label{ffac}
T_F(q_T^2,q^2)=e^{-B/3},\quad T_G(q_T^2,q^2)=e^{-3B/4},\quad
B=(\alpha_s/\pi) \ln^2 \left( \frac{q_T^2}{q^2}\right)\quad.
\end{equation}
These form factors make (render) the differential cross sections finite at
$q_T=0$. According to \cite{ellis81} they also
contain the virtual corrections. 

If we assume, that the partons inside the proton contain an intrinsic
transverse momentum which is distributed according to a Gaussian
${\rm d}N/{\rm d}k^2_T\propto \exp{k_T^2/\sigma^2}$ with $\langle
k_T \rangle \sim 
600 \,{\rm MeV}$ then the
theoretical predictions describe the data well \cite{field}. 
This value is perhaps somewhat high, as 
deep inelastic lepton-nucleon scattering experiments suggest a smaller value
of about $\langle k_T \rangle \sim 400\,{\rm MeV}$.

Actually, the double-logarithmic approximation (DLA) does not represent the
``state of the art''. If one incorporates sub-leading terms 
\cite{col85} then primordial momentum is not needed to describe
the data. This implies that 
the contribution from the Drell-Yan graph vanishes
because ${\rm d}\sigma/{\rm d}q_T=0$ at $q_T=0$ and 
the resulting shape of the ${\cal O}(\alpha_s)$ contributions is rather changed.
Such an approach would go beyond the present calculations
and would be required if our calculations suggested a statistical
accuracy to be expected from the planned RHIC experiments (which they
do not).
Thus, we restrict 
ourselves to the DLA for the calculation of the event rates although
strictly speaking it is not valid for all values of $q_T$ \cite{col85}.

After having clarified the situation for the unpolarized cross section
we want to consider polarized scattering now. Transverse spin
asymmetries manifest themselves in a nontrivial distribution of the
azimuth of the outgoing leptons (see  
(\ref{asydy})). Their origin can be easily understood. The axis of the
incoming momenta in the cm frame and the spin axis define one
plane in space. The orientation of the second one, 
defined by the momenta of the
outgoing leptons and the ones of the entering partons, is therefore
not arbitrary.

\begin{figure}[hbt]
\begin{center}
\epsfysize=6cm
\leavevmode\epsffile{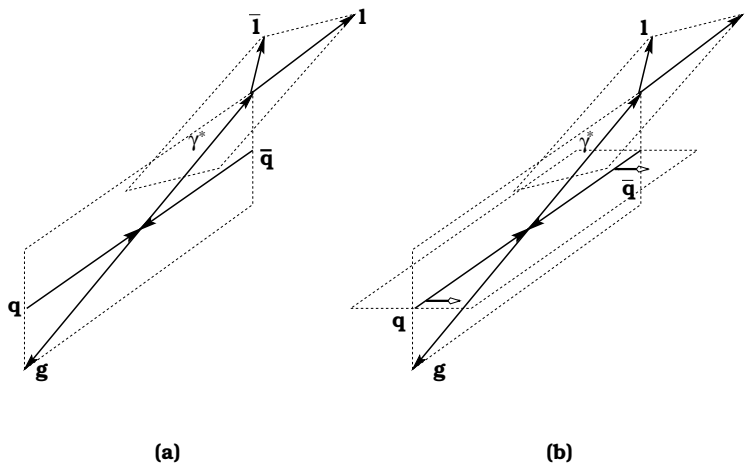}
\caption{\label{fig2}(a) the unpolarized QCD-annihilation process,
(b) the case with transverse polarization.}
\end{center}
\end{figure}

Figure~\ref{fig2}a shows the situation of the 
unpolarized QCD-annihilation graph which is similar to the previous one.
The orientation of the second plane in relation to that of the first one
is not arbitrary again. The major difference to the previous case is that the
first plane, defined by the momenta of the incoming partons and the gluon, 
now is not fixed in space because the azimuth of the gluon momentum
is not fixed. The same situation also holds for the QCD--Compton process.

Unfortunately, 
the calculation of the resulting angular distributions in QCD is not well
understood. One can try to use only perturbation theory
although $q_T$ becomes very small in the region we are interested in,
or one can invoke
resummation techniques for the small $q_T$ region but neither approach
describes the data completely \cite{chia}.

In case the incoming quark and antiquark are transversely polarized,
an additional third plane can be defined according to figure~\ref{fig2}b.
Here, the orientations of all planes in relation to each other are 
nontrivial. The virtual photon and the gluon now take on similar 
roles as the lepton pair in the Drell-Yan graph since they are the
outgoing particles from the first interaction. This means, that the
distribution of the azimuthal angle between the photon momentum and 
the spin axis shows the typical $\cos 2\phi$ modulation.

If one of the entering particles is a gluon then 
the azimuthal angle between the
first and the third plane is distributed evenly in the allowed interval, 
since gluons cannot carry an asymmetry. For this reason, the Compton graph
does not carry a transverse spin asymmetry at all. 

We did not make an attempt to calculate the transverse spin asymmetry of
the QCD-annihilation process because, as we mentioned above,
the angular distributions in the unpolarized case are not well understood.
Thus, one might expect to run into similar difficulties in the
polarized case.

For practical purposes we therefore made the crude but common 
assumption, that the
transverse asymmetry of the QCD-annihilation graph equals the 
Drell-Yan one.

\section{{\sc Sphinx tt}\label{3}}
In order to estimate the transverse double spin asymmetry in the framework
of our model we developped a new MC program for transversely polarized
nucleon-nucleon scattering called {\sc Sphinx tt}. It is based on the
Lund Monte Carlo {\sc Pythia~5.6} \cite{sjoe} and is similar 
to {\sc Sphinx}, a MC code for longitudinal polarized
scattering.

Important features of {\sc Sphinx tt} are:
\begin{itemize}
\item Polarization is described correctly in the initial state up to
      the hard partonic reaction. The spin orientations are not
      followed through the final states. String breaking and
      hadronization in general are assumed not to depend on the spin
      orientation.

\item Two sets of polarized parton distributions are available by now.
      The parametrizations of $\Delta q(x,Q^2)$
      by Gehrmann/Stirling \cite{gehr95} and Bartelski/Tatur \cite{bar95} 
      are taken as transversity distributions at $Q_0^2=4{\rm GeV^2}$ 
      and are evoluted according to the GLAP-equation (\ref{glap}).
      The implementation of other model parametrizations is easy.

\item Transverse polarization is correctly taken care of
      in the initial state showering
      routine. 

\item The polarized partonic cross sections for
      high-$p_T$ jet production, prompt photon production and
      lepton pair production in the Drell-Yan model are available.
      $Z^0$ physics is not implemented yet.

\item The majority of options of {\sc Pythia 5.6} is also available for
      {\sc Sphinx~tt}.
\end{itemize}

\section{Results\label{4}}
Here we present the results of our MC calculations of $A^{TT}$
for the PHENIX detector at RHIC. 

Our results are presented for $\sqrt s=50\,{\rm GeV}$ and an integrated
luminosity of ${\cal L}=200{\rm pb}^{-1}$ as well as $\sqrt s=200\,{\rm GeV}$
and ${\cal L}=800{\rm pb}^{-1}$. This is equivalent to 
230 days of beam time with 50\% efficiency.
The assumed integrated luminosity is larger by a factor of 2.5
compared to the standard value as quoted in \cite{rsc92, rsc93}
but is necessary in order to detect an adequate number of events \cite{rsc94}.  

Taking into account this larger integrated luminosity as well as the
expected Drell-Yan event rates for the STAR detector 
\cite{rsc93} the situation for this second detector is the same 
as presented below.

We assume that both endcaps will have a polar opening angle of 
$10^{\circ}-35^{\circ}$ with respect to the beam axis 
and will be able to measure myons with
$1.1 < y^{\mu} < 2.5$. 
The central muon detectors are not very well suited
for the measurement of $A^{TT}$ since they do not cover
the full azimuth.

MC calculations \cite{rsc94} indicate competition from charm decay for
$M_{\mu^+\mu^-} < 4\,{\rm GeV}$ at $\sqrt s=50\,{\rm GeV}$ and 
$M_{\mu^+\mu^-}  < 5\,{\rm GeV}$ at $\sqrt s=200\,{\rm GeV}$ so that we 
set the minimum lepton pair mass accordingly. We did not set a cut
to exclude the $\Upsilon$ production region at $8\,{\rm GeV} < 
M_{\mu^+\mu^-}  < 10\,{\rm GeV}$ because the cross section
decreases exponentially with increasing $M_{\mu^+\mu^-}$ anyway.

\begin{figure}[hbt]
\begin{center}
\epsfysize=16cm
\leavevmode\epsffile{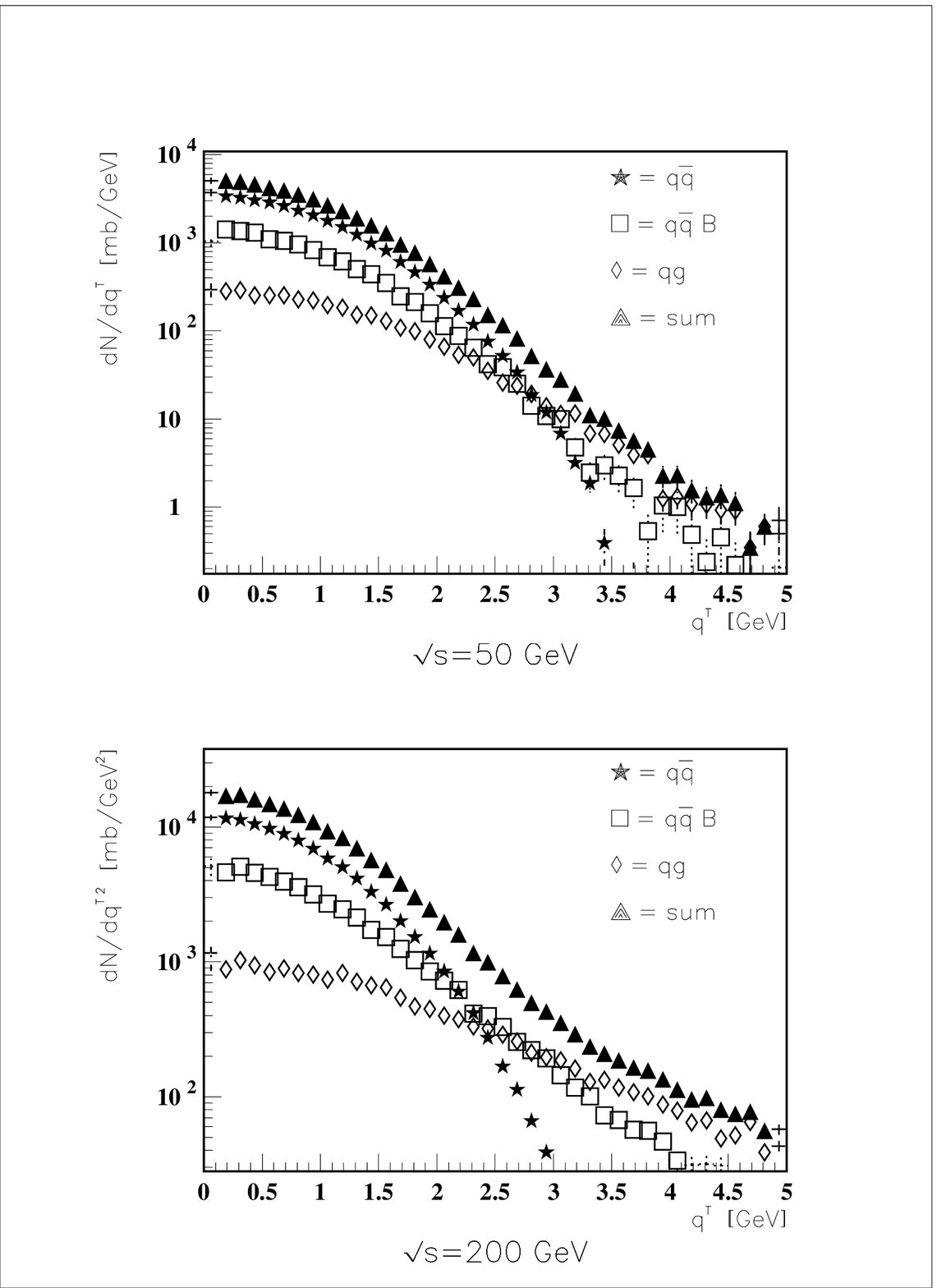}
\caption{\label{fig3} Distribution of the the transverse
momentum of the virtual photon $q_T$ at $\sqrt s=50\,{\rm GeV}$ and 
$\sqrt s=200\,{\rm GeV}$. $q\bar q$ denotes the Drell-Yan graph, $q\bar q$~$B$ the
QCD annihilation graph (B stands for Bremsstahlung).}  
\end{center}
\end{figure}

Figure~\ref{fig3} shows the differential cross section 
${\rm d}\sigma/{\rm d}q_T^2$ for both cm energies. The
Drell-Yan graph obviously dominates lepton pair production for
low $q_T$. The QCD Compton graph is the least important reaction in
this region.
In the high $q_T$ region we have the opposite situation.
To enhance the Drell-Yan signal
we therefore
required $q_T < 2\,{\rm GeV}$. 
This leads to an improvement of the
$K^{DY}$-factor by $\Delta K^{DY}=-0.1$ for $\sqrt s=50\,{\rm GeV}$ and
$\Delta K^{DY}=-0.3$ for $\sqrt s=200\,{\rm GeV}$ in our model. 
We multiplied the usual
${\cal O}(\alpha_s)$-matrix elements of the hard partonic scattering with the 
corresponding form factors~(\ref{ffac}). Afterwards we fitted the
width of the intrinsic momentum to the data at $\sqrt s=62\,{\rm GeV}$
taken from 
\cite{ant81}. Unfortunately, we got $\langle k_T \rangle =800\, {\rm MeV}$
which is very high in comparison to DIS experiments. On the other hand
we checked that the asymmetry does not depend on $\langle k_T \rangle$
for values
smaller than 1~GeV. Our results are rather independent of the details.
One should also note, that we switched off the general initial state
showering routine 
because it is only suited for the large $q_T$ region and to avoid
double counting. Bremsstrahlung effects are only taken into account
by the QCD-annihilation and Compton process.

\begin{figure}[hbt]
\begin{center}
\epsfysize=16cm
\leavevmode\epsffile{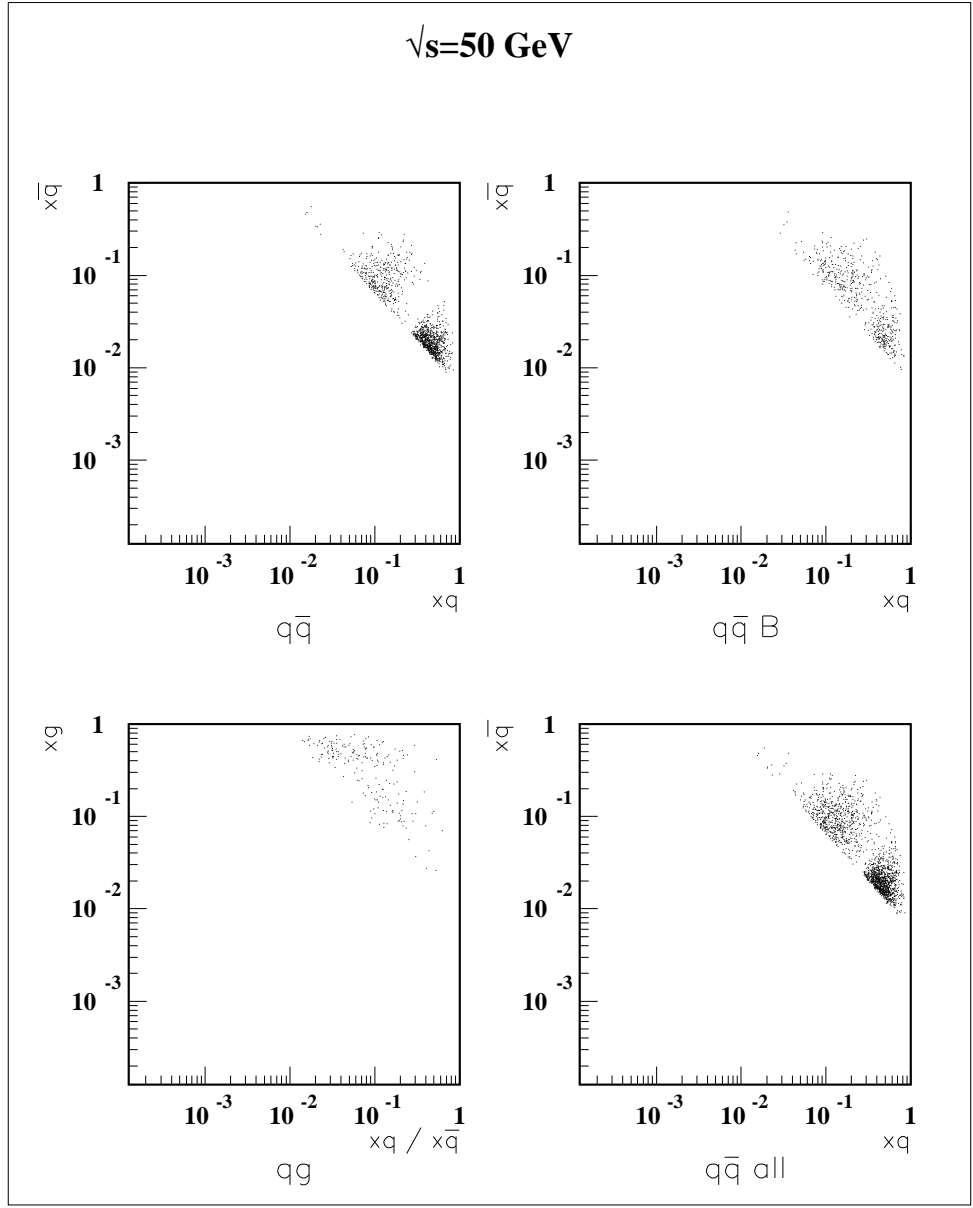}
\caption{\label{fig4} Event distribution in the plane of the 
Bj\o rken-$x$ of the partons that participate in the hard reaction
at $\sqrt{s}=50$ GeV.
$q\bar q$ denotes the Drell-Yan graph, $q\bar q$~$B$ the
QCD annihilation graph.}
\end{center}
\end{figure}

\begin{figure}[hbt]
\begin{center}
\epsfysize=16cm
\leavevmode\epsffile{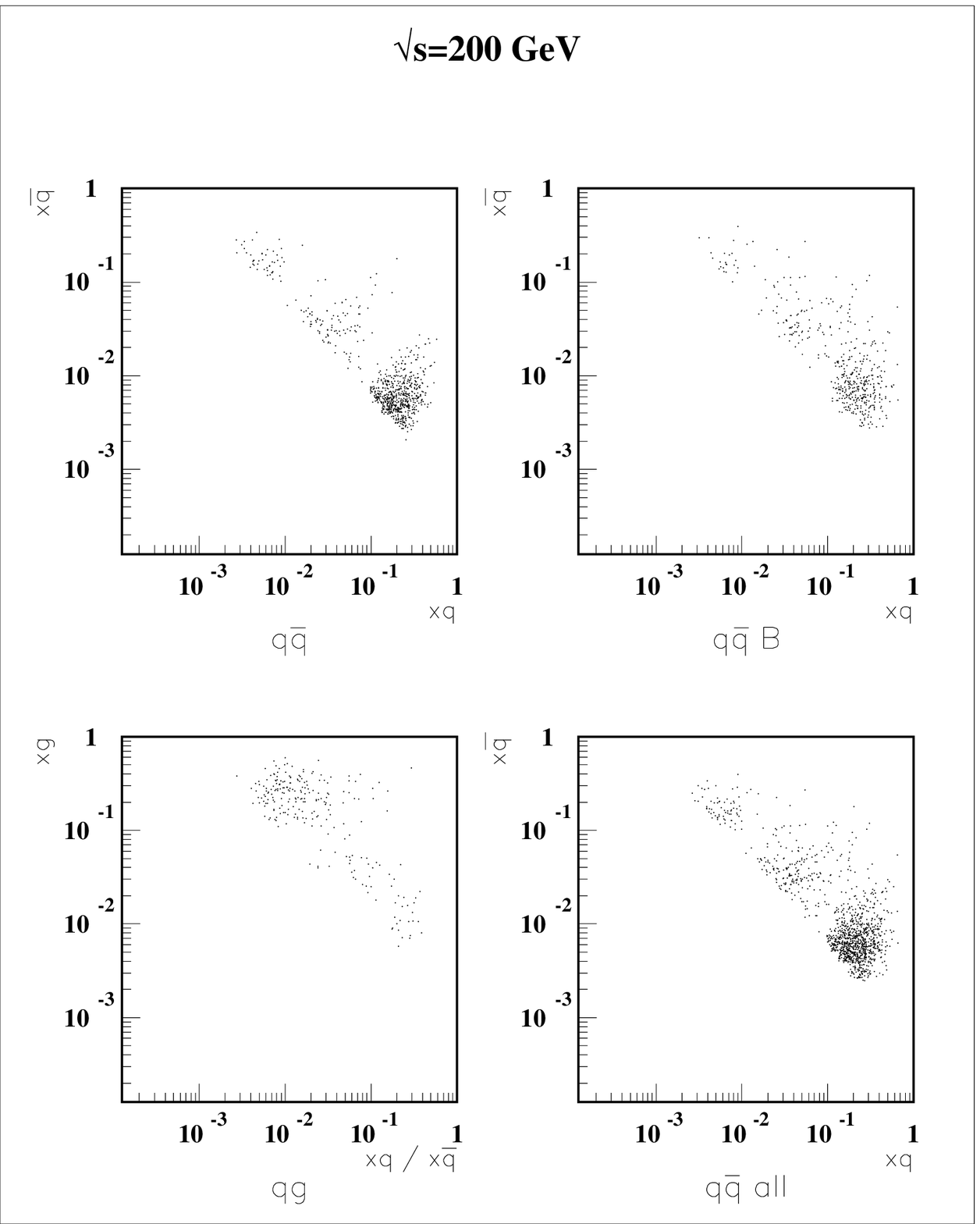}
\caption{\label{fig5}
Same as figure~\ref{fig4} for $\sqrt s=200$~GeV.}
\end{center}
\end{figure}

Figures~\ref{fig4} and \ref{fig5} show the distribution of events in the
plane of the Bj\o rken variables of the partons which participate in the hard 
interaction. Generally, the $x$-variables at $\sqrt s=200\, {\rm GeV}$ 
have smaller values as the ones at $\sqrt s=50\,{\rm GeV}$. In both cases,
there is one region in the $x$-$x$-plane where most of the Drell-Yan type 
events are situated. These regions can be characterized by
$0.3<x_q<0.7$ and $0.01<x_{\bar q}<0.04$ at $\sqrt s=50\,{\rm GeV}$, and
$0.1<x_q<0.4$ and $0.002<x_{\bar q}<0.02$ at $\sqrt s=200\,{\rm GeV}$
respectively.

Fortunately, these ranges are also the ones, where the largest asymmetries
are expected since the polaized parton distributions multiplied with $x$ have
a maximum there. Therefore,
it would be interesting to restrict the detected events to these regions.
Furthermore, the exchange terms in equation~(\ref{asydy}) would
disappear.
In order to achieve this one has to calculate $x_{A/B}$ from the
lepton momenta. Then one must 
decide which one of the incoming partons was the quark and which one
was the antiquark. 
Since the majority of events fulfill 
$x_q > x_{\bar q}$ this criterium can be used very successfully 
for this purpose. In only 2\%-10\% of all cases the $x$-values are
not assigned correctly. 

Of course, it is not possible to reconstruct the $x$-values of the
incoming partons of the QCD processes because the outgoing
quark or gluon jet is not detected. Since the jet momentum is quite
small in general we don't make a big error by using the Drell-Yan 
formulas for the reconstruction of the Bj\o rken-$x$ 
also for the QCD processes. 

Table~\ref{tab1} shows the event rates for the different cases.
At $\sqrt s=200\, {\rm GeV}$ they are about ten to twenty
times higher than for $\sqrt s=50\,{\rm GeV}$. This is mainly due to the 
fact that the unpolarized parton distributions rise steeply for smaller
$x$. However, according to the shape of the transversity distributions 
for $\sqrt s=200\,{\rm GeV}$ one should expect a smaller $A^{TT}$.
So we should try to minimize the statistical errors by
integrating over all kinematical variables.
The integration over the azimuth is performed as follows
\cite{cortes91}
\begin{equation}
A^{TT}=\left\{\int\limits_{-\frac{\pi}{4}}^{\frac{\pi}{4}} -
\int\limits_{\frac{\pi}{4}}^{\frac{3\pi}{4}}
+\int\limits_{\frac{3\pi}{4}}^{\frac{5\pi}{4}}
-\int\limits_{5\frac{\pi}{4}}^{7\frac{\pi}{4}}
\right\} {\rm d} \phi \,A^{TT}(\phi)\quad .
\end{equation}
For $\cos 2\phi$ this gives simply a factor of 4.
According to our model the QCD annihilation graph carries the same
asymmetry as the Drell-Yan graph which leads to
\begin{equation}
A^{TT}=\frac{N_{\rm DY}+N_{\rm annih.}}{N_{\rm DY}+N_{\rm annih.}+
N_{\rm Compton}} A^{TT}_{\rm DY}\quad .
\end{equation}

The statistical error of the asymmetry is $1/\sqrt N$ where
$N$ denotes the total event rate.

Table~\ref{tab2} contains the results for the asymmetry together with the
absolute statistical error.
We can summarize them as follows:
\begin{itemize}
\item If $\delta q(x,4\, {\rm GeV}^2)=\Delta q(x,4\,{\rm GeV}^2)$ then
$A^{TT}$ won't be measurable at RHIC
at $\sqrt s=50\,{\rm GeV}$ because the statistical error is
approximately as large as the asymmetry itself.

\item At $\sqrt s=200\,{\rm GeV}$ the {\em relative} statistical error is
also very large, namely $60\%-80\%$. Therefore, it is very unlikely that
the asymmetry can be measured as there will certainly be some unavoidable
background and systematic error.

\item The main reason for the predicted smallness of the
asymmetry is the very small antiquark polarization which is of the
order of 5\% according to common parametrizations of parton
distributions. By assuming a much larger sea polarization one gets 
larger asymmetries \cite{sof95}. Consequently, 
there is a good chance that $A^{TT}$ can be measured at
$\sqrt s=200\,{\rm GeV}$ if, e.g.,
$\delta \bar q(x,4\,{\rm GeV}^2)=3\Delta \bar q(x,4\,{\rm GeV}^2)$.

\item It is crucial for the experiment that RHIC will be able run with
the design luminosity in order to have enough events. 
\end{itemize}

When we were writing this article we received \cite{bar96}
which describes a confinement model calculation of $\delta q(x,Q^2)$.
The authors also calculate $A^{TT}$ and their results are of
the same order as the results presented in this paper.

\vspace{1cm}
{\bf Acknowledgements}\\
\\
We would like to thank Thomas Gehrmann for providing an evolution program
for helicity distributions. We are also grateful to
Torbj\"orn Sj\"ostrand for answering questions related to his program.
This work was supported by DFG~(Scha458/3) and BMBF.

\begin{table}
\begin{center}
\begin{tabular}{|l||c|c|}
\hline
\multicolumn{3}{|c|}{\rule[-3mm]{0mm}{8mm}\bf PHENIX}\\
\hline
\rule[-3mm]{0mm}{8mm}
subprocess & $\sqrt s=50\,{\rm GeV}$ & $\sqrt s=200\,{\rm GeV}$ \\
\hline
$q \bar q \rightarrow \mu^+\mu^- $& 5236&70390\\
$q \bar q \rightarrow \mu^+\mu^- g$&2157 &34720\\
$q g \rightarrow \mu^+\mu^- q $& 637&10770\\
\hline
sum $N$& 8030 & 115880 \\
$K$-factor & 1.5 & 1.6 \\
\hline
\hline
\multicolumn{3}{|c|}{\rule[-3mm]{0mm}{8mm}\bf PHENIX, $x$-$x$-Cuts}\\
\hline
\rule[-3mm]{0mm}{8mm}
subprocess & $\sqrt s=50\,{\rm GeV}$ & $\sqrt s=200\,{\rm GeV}$ \\
\hline
$q \bar q \rightarrow \mu^+\mu^- $& 3302&58020\\
$q \bar q \rightarrow \mu^+\mu^- g$&1027 &27540\\
$q g \rightarrow \mu^+\mu^- q $& 353&8740\\
\hline
sum $N$& 4682 & 94300 \\
$K$-factor & 1.4 & 1.6 \\
\hline

\end{tabular}
\caption{\label{tab1} Event rates.}
\end{center}
\end{table}

\begin{table}
\begin{center}
\begin{tabular}{|l||c|c|}
\hline
\multicolumn{3}{|c|}{\rule[-3mm]{0mm}{8mm}\bf PHENIX}\\
\hline
\rule[-3mm]{0mm}{8mm} parton distribution &$\sqrt s=50\,{\rm GeV}$ & 
$\sqrt s=200\,{\rm GeV}$ \\
\hline
Germann/Stirling & $1.00\%\pm 1.1\%$ & $0.38\%\pm 0.30\%$ \\
Bartelski/Tatur  & $0.63\%\pm 1.1\%$ & $0.51\%\pm 0.30\%$ \\
\hline
\hline
\multicolumn{3}{|c|}{\rule[-3mm]{0mm}{8mm}\bf PHENIX, $x$-$x$-Cuts}\\
\hline
\rule[-3mm]{0mm}{8mm} parton distribution& $\sqrt s=50\,{\rm GeV}$ &
$\sqrt s=200\,{\rm GeV}$ \\
\hline
Germann/Stirling & $1.05\%\pm 1.4\%$&$0.45\%\pm 0.35\%$ \\
Bartelski/Tatur  & $1.70\%\pm 1.4\%$&$0.54\%\pm 0.35\%$ \\
\hline
\end{tabular}
\caption{\label{tab2} Transverse double spin asymmetry $A^{TT}$
with absolute statistical error.
The forseen polarization of beam and target at RHIC (70\%) 
is already taken into account.}
\end{center}
\end{table}

\end{document}